\documentclass[pra,onecolumn,floatfix,aps,superscriptaddress,showpacs,amsfonts]{revtex4}
\usepackage{graphicx}
\usepackage{array}
\usepackage{amsthm}
\usepackage{amsmath}
\usepackage{amssymb}
\usepackage{mathrsfs}

\begin{document}


\title{Coherent-induced state ordering with fixed mixedness}

\author{Fu-Gang Zhang}
\affiliation{School of Mathematics and Information Science, Shaanxi Normal University, Xi'an, 710119, China}
\affiliation{School of Mathematics and Statistics, HuangShan University, Huangshan, 245041, China}


\author{Yongming Li}
\email{liyongm@snnu.edu.cn}
\affiliation{School of Mathematics and Information Science, Shaanxi Normal University, Xi'an, 710119, China}
\affiliation{College of Computer Science, Shaanxi Normal University, Xi'an, 710119, China}

\date{\today}
\begin{abstract}
In this paper, we study coherence-induced state ordering with Tsallis relative entropy of coherence, relative entropy of coherence and $l_{1}$ norm of coherence. Firstly, we show that these measures give the same ordering for single-qubit states with a fixed mixedness or a fixed length along the direction $\sigma_{z}$.
Secondly, we consider some special cases of high dimensional states, we show that these measures generate the same ordering for the set of high dimensional pure states if
any  two states  of the set satisfy majorization relation. Moreover, these three measures generate the same ordering for all $X$ states with a fixed mixedness.  Finally, we discuss dynamics of coherence-induced state ordering under Markovian channels. We find phase damping channel don't change the coherence-induced state ordering for some single-qubit states with fixed mixedness, instead amplitude damping channel  change the coherence-induced ordering even though for single-qubit states with fixed mixedness.


\end{abstract}

\maketitle
\section{Introduction}\label{intro}
Quantum coherence is one of the most important physical resources in quantum mechanics, which
can be used in quantum  optics~\cite{Scully97}, quantum information and quantum computation~\cite{Nielsen}, thermodynamics~\cite{Rodr13,berg14},
and low temperature thermodynamics~\cite{Horodecki13,Lostaglio15,Naras15}. Many efforts have been made in quantifying the coherence of quantum states~\cite{berg06}. The authors of Ref.~\cite{Baum14} proposed a rigorous framework to quantify coherence. The framework gives four conditions that any proper measure of the coherence must satisfy. Based on this framework, one can define suitable measures with respect to the prescribed orthonormal basis.
The relative entropy of coherence and the $l_{1}$ norm of coherence~\cite{Baum14} have been proved to satisfy these conditions. Recently, the author of Ref.~\cite{Rastegin16} has proposed  Tsallis relative entropy of coherence. The author has proved Tsallis relative entropy of  coherence
satisfies the conditions of (C1),(C2a) and (C3), but violate the condition of (C2b), i.e. Monotonicity under incoherent selective measurements. Whereas, this coherence measure satisfy a generalized monotonicity for average coherence under subselection based on measurement ~\cite{Rastegin16}.  In addition, various other coherence measures have also been  discussed~\cite{Rastegin16,Swapan16,Yuan15,Shao15,Strel15,Napoli16,Zhang16,Yu16,Chin17}.
Many further discussions about quantum coherence have been aroused~\cite{Liu16,Zhangfu16,Yao15,Du15,Cheng15,Bera15,Xi15,Winter16,Bromley15,Xu16,Yadin15,Bagan15,Chitambar16,Peng16,Bran15,Liu17,Tan17,Hu17}.

Up to now, many different coherence measures have been proposed based on different physical contexts. For the same state, different values of coherence will be obtained by different coherence measures. In this case, a very important question arises, whether these measures generate
the same ordering.
We say that two coherence measures  $C_{m}$ and $C_{n}$ generate the same ordering  if they satisfy the condition $C_{m}(\rho)\leq C_{m}(\sigma)$ if and only if $C_{n}(\rho)\leq C_{n}(\sigma)$ for any density operators $\rho$ and $\sigma$. Ref~\cite{Liu16}
 and ~\cite{Zhangfu16} have showed that the Tsallis relative entropy of  coherence, relative entropy of coherence and the  $l_{1}$ norm of coherence only generate the same ordering for single-qubit pure states. They don't give the same ordering for single-qubit mixed states or high dimension states even though high dimension pure states.
Based on these discussions, some further questions will be put forward as follows. (1) In addition to single-qubit pure states, whether or not there exist  other sets  of states such that above  coherence measures generate the same ordering. (2) Whether or not quantum operator change coherence-induced state ordering.

In the paper, we will try to resolve these two questions. Our discussion focus on
the Tsallis relative entropy of  coherence, relative entropy of coherence and the  $l_{1}$ norm of coherence.
For question(1), we  show these three measures  generate the same ordering for some particular sets of states, such as for some single-qubit states with a fixed mixedness or a fixed length along the direction $\sigma_{z}$. For question(2), we discuss dynamics of coherence-induced state ordering under Markovian channels, we show phase damping channel won't change the coherence-induced ordering for some single-qubit states with fixed mixedness, but amplitude damping channel change the coherence-induced ordering even though for single-qubit states with fixed mixedness. Other  Markovian channels can be discussed by a similar method.

This paper  is organized as follows.  In Sec.~\ref{sec:pre}, we briefly review some notions related to Tsallis relative entropy of coherence, relative entropy of coherence and $l_{1}$ norm of coherence.  In Sec.~\ref{sec:single}, we show that  Tsallis relative entropy of coherence, relative entropy of coherence and $l_{1}$ norm of coherence generate the same ordering for single-qubit states with a fixed mixedness or a fixed length  along the direction $\sigma_{z}$. In Sec.~\ref{sec:high}, we show that they  generate the same ordering for some particular sets of high dimensional states.
In Sec.~\ref{sec:dy}, we discuss dynamics of coherence-induced ordering under Markovian channels.
 We summarize our results in Sec.~\ref{sec:conclusion}.

\section{Preliminaries}\label{sec:pre}


In this section, we review some notions related to quantifying quantum coherence.
Considering a finite-dimensional Hilbert space $H$ with $d=dim(H)$.
Let $\{|i\rangle, i=1,2,\cdots,d\}$ be a particular basis of $H$. A state is called an incoherent state if and only if its density operator is
diagonal in this basis, and the set of all the incoherent states is usually denoted as $I$.
Baumgratz et al.~\cite{Baum14} proposed that $C$ is
a measure of quantum coherence if it satisfies following
properties: $(C1)$ $C(\rho)\geq 0$ and $C(\rho)= 0$ if and only if $\rho\in I$; $(C2a)$ $C(\rho)\geq C(\Phi(\rho))$, where $\Phi$ is any incoherent completely positive and trace preserving maps; $(C2b)$ $C(\rho)\geq \sum_{i}p_{i}C(\rho_{i})$, where $p_{i}=Tr(K_{i}\rho K^{\dag}_{i})$, $\rho_{i}=\frac{K_{i}\rho K^{\dag}_{i}}{Tr(K_{i}\rho K^{\dag}_{i})}$, for all ${K_{i}}$ with
$\sum_{i}K_{i}K^{\dag}_{i}=I$ and $K_{i}IK^{\dag}_{i}\subseteq I$; $(C3)$ $\sum_{i} p_{i}C(\rho_{i})\geq C(p_{i}\rho_{i})$ for any ensemble $\{p_{i},\rho_{i}\}$.

In accordance with the criterion, several coherence measures have been studied.
It has been  shown that  $l_{1}$ norm of coherence  and relative entropy of coherence satisfy
these four conditions ~\cite{Baum14}.
$l_{1}$ norm of coherence~\cite{Baum14} is defined as

\begin{equation}\label{l1norm}
C_{l_{1}}(\rho)=\displaystyle\sum_{i\neq j}\mid \rho_{ij}\mid,
\end{equation}
here $\rho_{ij}$ are entries of $\rho$. The coherence measure defined
by the $l_{1}$ norm is based on the minimal distance of $\rho$
to the set of incoherent states $I$, $C_{D}(\rho) = min_{\delta\in I} D(\rho,\delta)$
with D being the $l_{1}$ norm, and $0 \leq C_{l_{1}}(\rho) \leq d-1$.

The relative entropy of coherence~\cite{Baum14} is defined as
\begin{equation}\label{rel}
C_{r}(\rho)=\displaystyle\min_{\delta\in I}S(\rho||\sigma)=S(\rho_{diag})-S(\rho),
\end{equation}
where $S(\rho||\sigma) = Tr(\rho log_{2}\rho-\rho log_{2}\sigma)$ is the quantum relative entropy,
 $S(\rho) = Tr(\rho log_{2}\rho)$ is the von Neumann entropy, and $\rho_{diag}=\sum_{i}|i\rangle\langle i|\rho|i\rangle\langle i|$. The coherence measure defined by the
relative entropy is based on the minimal distance of $\rho$ to
$I$, $C_{D}(\rho) = min_{\delta\in I} D(\rho,\delta)$ with $D$ being the relative
entropy, and $0 \leq C_{r}(\rho) \leq log_{2} d$.

Tsallis relative entropy of coherence~\cite{Rastegin16}, denoted by $C_{\alpha}(\rho)$, is defined as
\begin{equation}
C_{\alpha}(\rho)=\displaystyle\min_{\delta\in I}D_{\alpha}(\rho\|\delta), \nonumber
\end{equation}
for any $\alpha\in(0,1)\sqcup (1,2]$. $D_{\alpha}(\rho\|\delta)$ is
Tsallis relative entropy~\cite{Furuichi04,Hiai11,Luo16} for the density matrices $\rho$ and $\delta$, which is defined as

\begin{equation}
D_{\alpha}(\rho\|\delta)=\frac{Tr(\rho^{\alpha}\delta^{1-\alpha})-1}{\alpha-1}, \nonumber
\end{equation}
for $\alpha\in(0,1)\sqcup (1,\infty)$. $D_{\alpha}(\rho\|\delta)$ reduces to the von Neumann relative entropy when $\alpha\rightarrow1$~\cite{Furuichi04}, i.e.,
$\displaystyle\lim_{\alpha\rightarrow 1} D_{\alpha}(\rho\|\delta)=S(\rho\|\delta)=Tr[\rho(\ln\rho-\ln\delta)]$.
Therefore, $C_{\alpha}(\rho)$ reduces to relative entropy of coherence $C_{r}(\rho)$ when $\alpha\rightarrow1 $~\cite{Rastegin16}.

The author of Ref.~\cite{Rastegin16}  proved that $C_{\alpha}$ satisfies the conditions of (C1), (C2a) and (C3) for all $\alpha\in(0,2]$, but it violates (C2b) in some situations. However, $C_{\alpha}$ satisfies a generalized monotonicity for the average coherence under subselection based on measurement as the following form~\cite{Rastegin16}.  Tsallis relative $\alpha$-entropy of coherence $C_{\alpha}(\rho)$ satisfies

\begin{equation}\label{C2b'}
\sum_{i}p_{i}^{\alpha}q_{i}^{1-\alpha}C_{\alpha}(\rho_{i})\leq C_{\alpha}(\rho),
\end{equation}
 where $\alpha\in(0,1)\cup(2]$, $p_{i}=Tr(K_{i}\rho K^{\dag}_{i})$, $q_{i}=Tr(K_{i}\delta_{\rho} K^{\dag}_{i})$,
and $\rho_{i}=\frac{K_{i}\rho K^{\dag}_{i}}{p_{i}}$.

A .E. Rastegin  gave an elegant mathematical analytical  expression of Tsallis relative $\alpha$-entropy of coherence~\cite{Rastegin16}.
For all $\alpha\in(0,1)\cup(1,2]$, given a state $\rho$, the  Tsallis relative $\alpha$-entropy of coherence $C_{\alpha}(\rho)$ can be expressed as

\begin{equation}\label{alpha}
C_{\alpha}(\rho)=\frac{r^{\alpha}-1}{\alpha-1},
\end{equation}
where $r=\sum_{i}\langle i|\rho^{\alpha}|i\rangle^{\frac{1}{\alpha}}$. For the given $\rho$ and $\alpha$, based on this coherence measure, the nearest incoherence state from $\rho$ is the state

\begin{equation}
\delta_{\rho}=\frac{1}{r}\sum_{i}\langle i|\rho^\alpha|i\rangle^\frac{1}{\alpha}|i\rangle\langle i|. \nonumber
\end{equation}

 Considering an interesting case $\alpha=2$

\begin{equation}
C_{2}(\rho)=\Big(\sum_{j}\sqrt{\sum_{i}|\rho_{i,j}|^{2}}\Big)^2-1,
\end{equation}
where $\rho_{i,j}=\langle i|\rho|j\rangle$. $C_{2}$ is a function of squared module $|\rho_{i,j}|^{2}$, we should  distinguish it from squared $l_{2}$ norm of coherence $C_{l_{2}}$, where
$C_{l_{2}}$ is defined as

\begin{equation}
C_{l_{2}}(\rho)=\displaystyle\sum_{i\neq j}\mid \rho_{ij}\mid^2. \nonumber
\end{equation}
It has been shown that  $C_{l_{2}}$ doesn't satisfy the condition (C2b)~\cite{Baum14}. Although $C_{2}$ also violates the condition (C2b), it obeys a generalized monotonicity property Eq. (\ref{C2b'})~\cite{Rastegin16}.

We say the state  $\rho$ is a pure state if $Tr(\rho^2) = Tr(\rho) = 1$. If $\rho$ is not pure, then we say it is a mixed state.  For an arbitrary d-dimensional state, the mixedness
based on normalized linear entropy ~\cite{Peters04} is given as

\begin{equation}\label{mixedness}
M(\rho)=\frac{d}{d-1}(1-Tr(\rho^2)).
\end{equation}
In particular, when $\rho$ is a single-qubit state, $M(\rho)=2(1-Tr(\rho^2))$.

\section{Single-qubit states}\label{sec:single}

We first consider 2-dimensional quantum systems. A general single-qubit state can be written as

 \begin{equation}\label{equa}
\rho=\frac{1}{2}(I+\vec{k}\vec{\sigma}),
\end{equation}
 where $\vec{k}=(k_{x},k_{y},k_{z})$ is a real vector satisfying $\parallel\vec{k}\parallel\leq 1$, and $\vec{\sigma}=(\sigma_{x},\sigma_{y},\sigma_{z})$ is the vector of Pauli matrices. Any single-qubit state can be characterized by a vector  $\vec{k}$.
 In order to present our results, we classify the states with the form Eq. (\ref{equa}). Let $t=\parallel\vec{k}\parallel\leq 1$, and $\vec{n}=(n_{x},n_{y},n_{z})=\frac{1}{t}\vec{k}$.
where $\vec{n}=(n_{x},n_{y},n_{z})$ is a unit vector,
$t$ represents the length of vector $\vec{k}$, and $n_{x}(n_{y},n_{z})$ represents the length of vector $\vec{k}$ along the direction $\sigma_{x}(\sigma_{y},\sigma_{z})$.
By substituting  $\vec{k}=t\vec{n}$ into Eq. (\ref{equa}), we have

\begin{equation}\label{equa1}
\rho=\frac{1}{2}(I+t\vec{n}\vec{\sigma}).
\end{equation}

By substituting Eq. (\ref{equa1}) into Eq. (\ref{mixedness}),  we obtain the mixedness of single-qubit state $\rho$ as follow

\begin{equation}\label{mixedness1}
M(\rho)=2(1-tr(\rho^2))=1-t^2.
\end{equation}
According to the expression of  mixedness, we find that the mixedness only relates to the length $t$,
and it doesn't relate to the vector $\vec{n}=(n_{x},n_{y},n_{z})$. When $t=1$, the state becomes a special pure state. In Ref~\cite{Liu16}
and ~\cite{Zhangfu16}, authors find $C_{l_{1}}$, $C_{r}$ and $C_{\alpha}$ give the same ordering states for all single-qubit pure states. Here, we first  generalize this result. We will show that these coherence measures also give the same ordering for all states with a fixed mixedness.

With an easy calculation, we obtain the eigenvalues of  $\rho$,

\begin{equation}\label{eigenvalues}
\lambda_{1}=\frac{1+t}{2}, \lambda_{2}=\frac{1-t}{2}.
\end{equation}
Their norm eigenvectors are

\begin{equation}\label{eigenvectors1}
|\lambda_{1}\rangle=[\sqrt{\frac{1+n_{3}}{2}},\sqrt{\frac{1-n_{3}}{2}}],
\end{equation}

\begin{equation}\label{eigenvectors2}
|\lambda_{2}\rangle=[\sqrt{\frac{1-n_{3}}{2}},\sqrt{\frac{1+n_{3}}{2}}],
\end{equation}
By substituting Eq. (\ref{equa1}), (\ref{eigenvalues}), (\ref{eigenvectors1}), (\ref{eigenvectors2}) into Eq. (\ref{l1norm}), (\ref{rel}), (\ref{alpha}), we obtain $l_{1}$-norm coherence measure
\begin{equation}\label{l1norm1}
C_{l_{1}}(\rho)=t.\sqrt{1-n_{3}^2},
\end{equation}
Relative entropy of coherence
\begin{equation}\label{rel1}
C_{r}(\rho)=h(\frac{1+tn_{3}}{2})-h(\frac{1+t}{2}),
\end{equation}
where $h(x)=-xlog(x)-(1-x)log(1-x)$,
Tsallis relative $\alpha$-entropies of coherence
\begin{equation}\label{alpha1}
C_{\alpha}(\rho)=\frac{r^{\alpha}-1}{\alpha-1},
\end{equation}
where
\begin{equation}\label{r}
\begin{aligned}
r=&[(\frac{1+t}{2})^\alpha\frac{1+n_{z}}{2}+(\frac{1-t}{2})^\alpha\frac{1-n_{z}}{2}]^\frac{1}{\alpha}
  +[(\frac{1+t}{2})^\alpha\frac{1-n_{z}}{2}+(\frac{1-t}{2})^\alpha\frac{1+n_{z}}{2}]^\frac{1}{\alpha}.
 \end{aligned}
\end{equation}

According to  Eq. (\ref{mixedness1}), we know that all states  with the  same mixedness  have the same length $t$. First, we give the following proposition.

\textbf{Proposition 1}: For a fixed value $t$, The Eq. (\ref{l1norm1}), (\ref{rel1}), (\ref{alpha1}) are
decreasing functions with respect to $n_{z}$.

Proof: It is clear that the Eq. (\ref{l1norm1}) is a decreasing function with respect to $n_{z}$
for a fixed valued $t$.
Since $\frac{\partial C_{r}(\rho)}{\partial n_{z}}=\frac{t}{2}log(\frac{1-tn_{z}}{1+tn_{z}})\leq0$, we have that $C_{r}(\rho)$ is a decreasing function with respect to $n_{z}$.

 We first consider the derivation of the expression of Eq. (\ref{r}), we let $m=\frac{1}{2}((\frac{1+t}{2})^\alpha+(\frac{1-t}{2})^\alpha)$ and $n=\frac{1}{2}((\frac{1+t}{2})^\alpha-(\frac{1-t}{2})^\alpha)$, it is clear $m\geq n\geq 0$, then we have

\begin{center}
$\frac{\partial r}{\partial n_{z}}=\frac{n}{\alpha}(m+nn_{z})^{\frac{1}{\alpha}-1}
-\frac{n}{\alpha}(m-nn_{z})^{\frac{1}{\alpha}-1}$
$\left\{
\begin{array}
{cc}> 0,& \quad   \alpha <1 , \\
<0, &\quad  \alpha >1.
\end{array}
\right. $
\end{center}
Substituting this inequation into Eq. (\ref{rel1}),  we have $C_{\alpha}$ is a decreasing function.

In the following, we discuss the ordering states for single- qubit states with a fixed mixedness.

\textbf{Theorem 1}: For all single-qubit states with a fixed mixedness $M$, the coherence measures $C_{l_{1}}$,  $C_{r}$ and $C_{\alpha}$ will have the same
 ordering, where $\alpha\in(0,1)\cup(1,2]$.

Proof: Let $\rho$, $\sigma$ be two single-qubit states with a fixed mixedness  $M$. It is clear $\rho$, $\sigma$  have the same value $t$ by means of Eq. (\ref{mixedness1}). According to Proposition 1, we have $C_{l_{1}}(\rho)\leq C_{{1}}(\sigma)\Longleftrightarrow C_{r}(\rho)\leq C_{r}(\sigma)\Longleftrightarrow C_{\alpha}(\rho)\leq C_{\alpha}(\sigma)$.

Theorem 1 gives a  sufficient condition that these three coherence measures have the same ordering.
According to the  Eq. (\ref{l1norm1}), (\ref{rel1}), (\ref{alpha1}), we will find $C_{l_{1}}$,  $C_{r}$ and $C_{\alpha}$ relate to $t$ and $n_{z}$. We have shown that these coherence measures have the same ordering for all states with a fixed $t$. But it is quite natural that we  will ask whether these coherence measures have the same ordering for all states with a fixed length $n_{z}$ along the direction $\sigma_{z}$.

Since it is very difficult to discuss the monotony of the  expressions of $C_{\alpha}$ for all parameters $\alpha\in(0,1)\cup(1,2]$. Here, we only discuss the situations
when $\alpha=2,\frac{1}{2}$ by analytical method. In fact, we find
the results are  also valid for other values $\alpha\in(0,1)\cup(1,2]$ by numerical method. In Fig.1, we discuss the monotony of  $C_{\alpha}$ with respect to $t$ when $\alpha=\frac{1}{4},\frac{3}{4},\frac{3}{2}$ for fixed $n_{z}=\frac{1}{4},\frac{1}{2},\frac{3}{4}$. We find the following result is valid in these situations. For the other situations, we can discuss them by a similar method.

\textbf{Proposition 2}: For any fixed valued $n_{z}$,  Eq. (\ref{l1norm1}), (\ref{rel1}), (\ref{alpha1})
are increasing functions with respect to $t$, where $\alpha=\frac{1}{2},2$.

proof: It is clear that $C_{l_{1}}(\rho)$ is an increasing function with respect to $t$.

Since $\frac{\partial C_{r}(\rho)}{\partial t\partial n_{z}}=\frac{1}{2}log(\frac{1-tn_{z}}{1+tn_{z}})-\frac{t}{1-t^2n_{z}^2}\leq 0$, then $\frac{\partial C_{r}(\rho)}{\partial t}=\frac{n_{z}}{2}log(\frac{1-tn_{z}}{1+tn_{z}})-\frac{1}{2}log(\frac{1-t}{1+t})\geq 0$. Therefore,  $C_{r}(\rho)$ is an increasing function.

In order to discuss the monotonicity of $C_{2}$ and $C_{\frac{1}{2}}$ with respect to $t$, we first consider the monotonicity of $r$ (Eq. (\ref{r})) with respect to $t$ when $\alpha=\frac{1}{2},2$.  For convenience, Let $p=\frac{1+t}{2}$, $q=\frac{1+n_{z}}{2}$. It is clear $\frac{1}{2}\leq p,q\leq 1$. Substituting $p,q$ into Eq. (\ref{r}), and by a routine calculation we have
\begin{equation}
\frac{\partial r}{\partial p}\mid_{\alpha=2}=\frac{pq-(1-p)(1-q)}{\sqrt{p^2q+(1-p)^2((1-q)}}+\frac{p(1-q)-(1-p)q}
{\sqrt{p^2(1-q)+(1-p)^2q}}.\nonumber
\end{equation}
If $p\geq q$, it is easy to find   $\frac{\partial r}{\partial p}\mid_{\alpha=2}\geq 0$.
If $p\leq q$, we consider
\begin{equation}
\begin{aligned}
&\frac{[pq-(1-p)(1-q)]^2}{p^2q+(1-p)^2((1-q)}-\frac{[p(1-q)-(1-p)q]^2}{p^2(1-q)+(1-p)^2q}
  =\frac{q(1-q)(2q-1)(2p-1)}{(p^2q+(1-p)^2((1-q))(p^2(1-q)+(1-p)^2q)}\geq 0,\nonumber
 \end{aligned}
\end{equation}
it follows that $\frac{\partial r}{\partial p}\mid_{\alpha=2}\geq 0$.

When $\alpha=\frac{1}{2}$, by a routine calculation, we have
\begin{equation}\nonumber
\begin{aligned}
\frac{\partial r}{\partial p}\mid_{\alpha=\frac{1}{2}}=&[p^\frac{1}{2}q+(1-p)^\frac{1}{2}(1-q)][p^\frac{-1}{2}q-(1-p)^\frac{-1}{2}(1-q)]\\
&+[p^\frac{1}{2}(1-q)+(1-p)^\frac{1}{2}q][p^\frac{-1}{2}(1-q)-(1-p)^\frac{-1}{2}q]\\
 =&\frac{2(1-2p)}{\sqrt{p(1-p)}}\leq 0.
 \end{aligned}
\end{equation}

Substituting the above results into  Eq. (14), we have $\frac{C_{\frac{1}{2}}}{\partial p}\geq 0$ and $\frac{\partial C_{2}}{\partial p}\geq 0$. Therefore, $C_{2}$ and $C_{\frac{1}{2}}$ are increasing functions with respect to $t$ for a fixed $n_{z}$.

On the basis of the above proposition, we discuss the ordering  states for all single-qubit states with a fixed $n_{z}$.

\textbf{Theorem 2}: For all single-qubit states with a fixed  $n_{z}$, the coherence measures $C_{l_{1}}$,  $C_{r}$ and $C_{\alpha}$ have the same ordering, where $\alpha=\frac{1}{2},2$.

The proof is easy based on the Proposition 2. Theorem 2 gives another  sufficient condition that these three coherence measures have the same ordering. For any two single-qubit states, if they have the same
length along the direction $\sigma_{z}$, then $C_{l_{1}}$,  $C_{r}$ and $C_{\alpha}( \alpha=\frac{1}{2},2)$ take the same ordering for these two states. In fact, we find above result is also valid for any $\alpha\in(0,1)\cup(1,2]$ by numerical method.

\begin{figure}[t]
\begin{tabular}{cc}
\begin{minipage}[t]{1.5in}
\includegraphics[scale=0.18]{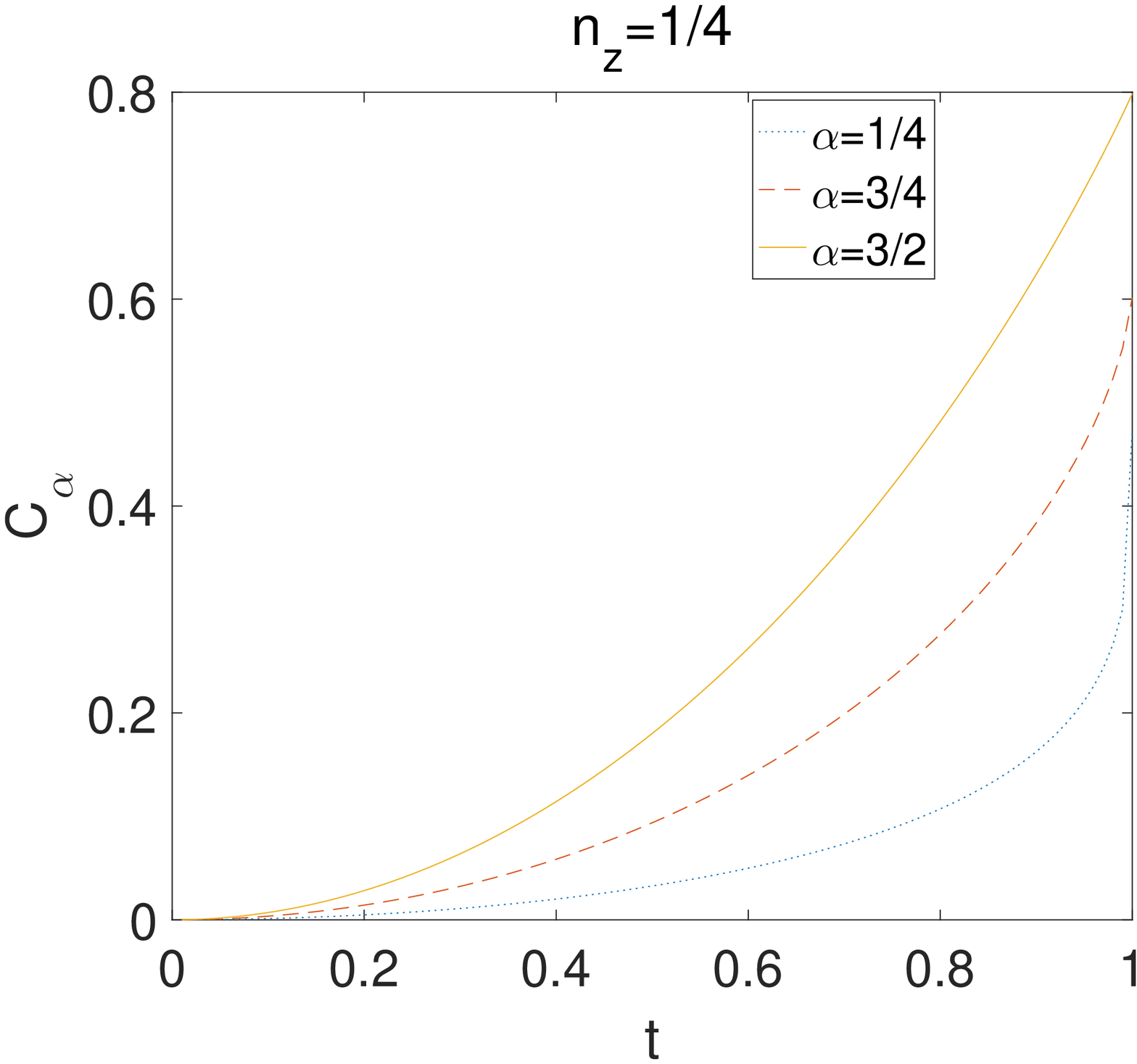}
\center
\end{minipage}
\begin{minipage}[t]{1.5in}
\includegraphics[scale=0.18]{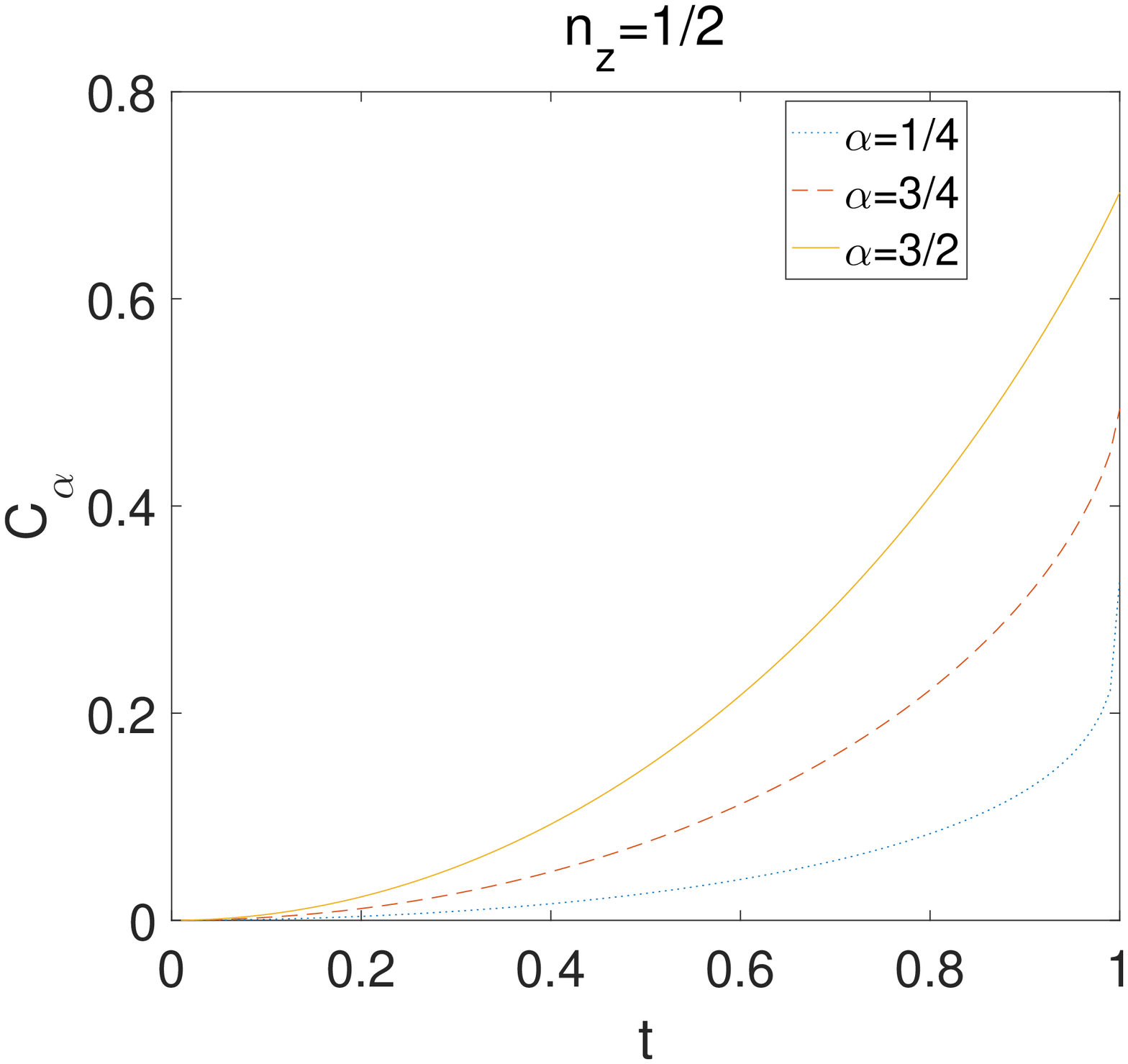}
\center
\end{minipage}
\begin{minipage}[t]{1.5in}
\includegraphics[scale=0.175]{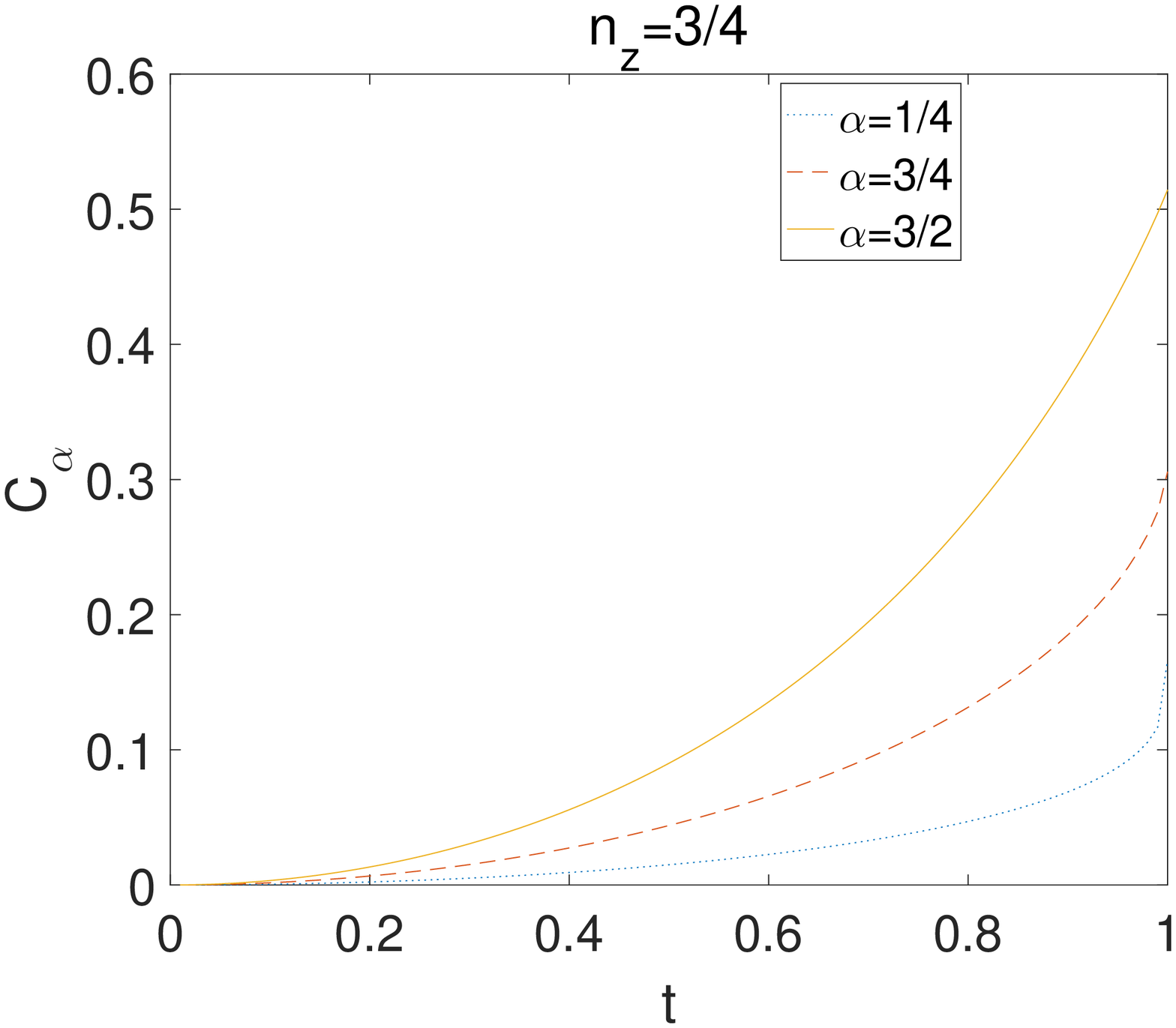}
\center
\end{minipage}
\end{tabular}
\center{Fig. 1. Three special Tsallis relative $\alpha$-entropies of coherence $C_{\frac{1}{4}}$, $C_{\frac{3}{4}}$ and $C_{\frac{3}{2}}$ are increasing functions with respect $t$ for the fixed $n_{z}=\frac{1}{4},\frac{1}{2},\frac{3}{4}$.}
\end{figure}

\section{ High-dimensional states}\label{sec:high}

In Ref~\cite{Liu16} and ~\cite{Zhangfu16}, it has been shown that these coherence measures don't generate the same ordering  for  high dimensional states, even though these states are pure. Here, we show  that they will generate the same ordering  when we restrict to some special states.

\subsection{ Pure states}

We first show  these coherence measures generate the same ordering for some special high-dimensional pure states.  We first introduce the notion  of shur-concave function~\cite{Mar79}.
For two vectors $\vec{x}$ and $\vec{y}$, we say that $\vec{x}$ is majorized by $\vec{y}$, denote it by x $\prec y$, if the rearrangement of
the components of $\vec{x}$ and $\vec{y}$, i.e., $x[1] \geq x[2] \geq \cdots \geq x[n]$, $y[1] \geq y[2] \geq \cdots \geq y[n]$, satisfies $\sum^{k}_{i=1}{x[i]}\leq \sum^{k}_{i=1}{y[i]}$, where $k\in[1,2,\cdots,n]$. For two vectors  $\vec{x}$ and $\vec{y}$, we say they satisfy majorization relation if $\vec{x} \prec \vec{y}$ or $\vec{y} \prec \vec{x}$.
The function $F : R^{n} \rightarrow R$, is called Schur-convex if $\vec{x} \prec \vec{y}$ implies $F(\vec{x}) \leq F(\vec{y})$. Function $F$ is called Schur-concave if -$F$ is Schur-convex.

\textbf{Lemma 1}~\cite{Mar79}: Let $F(\vec{x}) = F(x[1] \geq x[2] \geq \cdots \geq x[n])$ be a symmetric function with continuous partial
derivatives on $I^{n}$, where $I$ is an open interval. Then $F : I^{n}\rightarrow R$ is Schur convex if and only if the inequality
$(x_{i}-x_{j})(\frac{\partial F}{\partial x_{i}} - \frac{\partial F}{\partial x_{j}})\geq 0$
holds on  for each $i,j\in \{1,\cdots,n\}$. function $F$ is Schur-concave if the inequality  is reversed.

Given any $d$  dimensional pure state $|\psi\rangle=\displaystyle\sum_{i}\sqrt{\lambda_{i}}|i\rangle$, by means of Eq. (\ref{l1norm}), (\ref{rel}), (\ref{alpha}), we can obtain the $l_{1}-$ coherence, relative entropy of coherence and  Tsallis relative entropy of coherence as follows

\begin{equation}\label{diml1norm}
C_{l_{1}}(\psi)=(\displaystyle\sum_{i}\sqrt{\lambda_{i}})^2-1,
\end{equation}

\begin{equation}\label{dimrel}
C_{r}(\psi)=\displaystyle\sum_{i}(-\lambda_{i}log\lambda_{i}),
\end{equation}


\begin{equation}\label{dimalpha}
C_{\alpha}(\psi)=\displaystyle\sum_{i}\frac{r^\alpha-1}{\alpha-1}, where \quad r=\displaystyle\sum_{i}\lambda_{i}^\frac{1}{\alpha}.
\end{equation}


According to Lemma 1,  it is easy to show the following proposition.

\textbf{Proposition 3}: Eq. (\ref{diml1norm}), (\ref{dimrel}), (\ref{dimalpha}) are concave functions, where $\alpha\in(0,1)\cup(1,2]$.

According to the above proposition, we can easily obtain the following  theorem.

\textbf{Theorem 3}: Let $S$ be a set of d-dimensional pure states ($d\in Z^{+}$ and $d\geq 3$), if any two pure states $|\psi\rangle=\displaystyle\sum_{i}\sqrt{\lambda_{i}}|i\rangle,|\varphi\rangle
=\displaystyle\sum_{i}\sqrt{\mu}|i\rangle\in S$ satisfy majorization relation, then $C_{l_{1}}$,$C_{r}$ $C_{\alpha}$ have the same ordering for all states in $S$.

Theorem 3 gives a sufficient condition whether these coherence measures  generate the same ordering for some sets of d-dimensional pure states. But the following example will show that the inverse result don't hold. Two qutrit pure states are given as follows,
\begin{center}
 $\mid\psi_{1}\rangle=\frac{1}{2}\mid0\rangle+\frac{1}{4}\mid 1\rangle+\frac{1}{4}\mid 2\rangle$,

  $\mid\psi_{2}\rangle=\frac{2}{5}\mid0\rangle+\frac{2}{5}\mid 1\rangle+\frac{1}{5}\mid 2\rangle$.
\end{center}

  It is easy to calculate that $C_{l_{1}}(\mid \psi_{1}\rangle)= 1.9142$, $C_{l_{1}}(\mid\psi_{2}\rangle)= 1.9314$, $C_{r}(\mid\psi_{1}\rangle)=1.5000$, $C_{r}(\mid\psi_{2}\rangle)= 1.5219$, $C_{\frac{1}{2}}(\mid\psi_{1}\rangle)=0.7753$, $C_{\frac{1}{2}}(\mid\psi_{2}\rangle)= 0.8000$.
 So $C_{l_{1}}(\mid\psi_{1}\rangle)< C_{l_{1}}(\mid\psi_{2}\rangle)$, 
$C_{r}(\mid\psi_{1}\rangle)< C_{r}(\mid\psi_{2}\rangle)$ and
 $C_{\frac{1}{2}}(\mid\psi_{1}\rangle)< C_{}(\mid\psi_{2}\rangle)$.
Therefore, we  know that $C_{l_{1}}$ , $C_{r}$
and $C_{\frac{1}{2}}$
 generate the same  ordering for $\mid\psi_{1}\rangle$ and $\mid\psi_{2}\rangle$.
But $[\frac{1}{2},\frac{1}{4},\frac{1}{4}]$ and  $[\frac{2}{5},\frac{2}{5},\frac{1}{5}]$ don't satisfy majorization relation. As a result, we say the inverse of the Theorem 3 is invalid.

\subsection{ $X$ states}

Quantum states having "$X$"-structure are referred to as $X$ states. Consider an n-qubit $X$ state given by

\begin{equation}\label{xstate}
\rho=p|gGHZ\rangle \langle gGHZ|+(1-p)\frac{I_{d}}{d},
\end{equation}
where $|gGHZ\rangle=a|0\rangle^{\bigotimes n}+b|1\rangle^{\bigotimes n}$, with $a^2+b^2=1$, $I_{d}$
is an identity matrix, $d=2^{n}$ and $0\leq  p \leq 1$.

It is easy to calculate  that eigenvalues of $\rho$ are  $\lambda_{1}=p+\frac{1-p}{d}$, $\lambda_{2}=\lambda_{3},\cdots,=\lambda_{d}=\frac{1-p}{d}$, and their  eigenvectors  are $|\lambda_{1}\rangle=[|a|,0,\cdots,0,|b|]^T$, $|\lambda_{2}\rangle=[0,1,\cdots,0,0]^T$, $\cdots$, $|\lambda_{d-1}\rangle=[0,1,\cdots,0,0]^T$,$|\lambda_{d}\rangle=[|b|,0,\cdots,0,|a|]^T$.
By substituting  the eigenvalues and eigenvectors into the Eq. (\ref{l1norm}),(\ref{rel}),(\ref{alpha}), we have  $l_{1}$-norm coherence measure

\begin{equation}\label{xl1norm}
C_{l_{1}}(\rho)=2p|ab|.
\end{equation}
Relative entropy of coherence
\begin{equation}\label{xrel}
\begin{aligned}
C_{r}(\rho)&=S(\rho_{diag})-S(\rho)=-(p\cdot a^2+\frac{1-p}{d}) log(p\cdot a^2+\frac{1-p}{d})\\
&-(p\cdot b^2+\frac{1-p}{d})log(p\cdot b^2
+\frac{1-p}{d})
-(\frac{1-p}{d}) log(\frac{1-p}{d})-(p+\frac{1-p}{d}) log(p+\frac{1-p}{d}).
 \end{aligned}
\end{equation}
Tsallis relative $\alpha$-entropies of coherence
 \begin{equation}\label{xalpha}
 C_{\alpha}(\rho)=\frac{r^{\alpha}-1}{\alpha-1},
 \end{equation}
  where
 \begin{equation}\label{xr}
\begin{aligned}
 r=&[(p+\frac{1-p}{d})a^2+(\frac{1-p}{d})b^2]^\frac{1}{\alpha}
 +[(p+\frac{1-p}{d})b^2+(\frac{1-p}{d})a^2]^\frac{1}{\alpha}
 +(d-2)\frac{1-p}{d}.
  \end{aligned}
\end{equation}

Substituting $\rho$ into Eq. (\ref{mixedness}), we obtain the mixedness of the X state  $M(\rho) =p$. In the following, we will show that the result of  Theorem 1  is also valid for all X states  with the fixed mixedness.

\textbf{Proposition 4}: Eq. (\ref{xl1norm}), (\ref{xrel}), (\ref{xalpha}) have the same monotony with respect to $a$.

Since these coherence measures are symmetry with respect to $a=0.5$, so we let $a\in [0,\frac{1}{2}]$.
It is clear that $\frac{\partial C_{l_{1}}(\rho) }{\partial a}\geq 0$ according to Eq. (\ref{xl1norm}). We consider the derivation of $C_{r}$ with respect to $a$,

\begin{equation}\nonumber
 \frac{\partial C_{r}}{\partial a}=2pa\times log(\frac{pb^2+\frac{1-p}{d}}{pa^2+\frac{1-p}{d}})\geq 0,
\end{equation}
it follows that $C_{l_{1}}, C_{r}$ are decreasing functions with respect to $a$.

Before considering  the  monotony of $C_{\alpha}$(Eq. \ref{xr}), we first consider the  monotony of $r$ with respect to $a$.
 \begin{equation}\nonumber
\begin{aligned}
 \frac{\partial r}{\partial a}=&\frac{2a}{\alpha}[(p+\frac{1-p}{d})^\alpha- (\frac{1-p}{d})^\alpha]\cdot \{[(p+\frac{1-p}{d})^\alpha a^2+(\frac{1-p}{d})^\alpha b^2]^{\frac{1}{\alpha}-1}\\
 &-[(\frac{1-p}{d})^\alpha a^2+(p+\frac{1-p}{d})^\alpha b^2]^{\frac{1}{\alpha}-1}\}.
\end{aligned}
\end{equation}
When $\alpha\in (1,2]$. If $(p+\frac{1-p}{d})^\alpha\geq(\frac{1-p}{d})^\alpha$, then it is easy to show
$[(p+\frac{1-p}{d})^\alpha a^2+(\frac{1-p}{d})^\alpha b^2]^{\frac{1}{\alpha}-1}
 \geq[(\frac{1-p}{d})^\alpha a^2+(p+\frac{1-p}{d})^\alpha b^2]^{\frac{1}{\alpha}-1}$. If $(p+\frac{1-p}{d})^\alpha\leq(\frac{1-p}{d})^\alpha$, then $[(p+\frac{1-p}{d})^\alpha a^2+(\frac{1-p}{d})^\alpha b^2]^{\frac{1}{\alpha}-1}
 \leq[(\frac{1-p}{d})^\alpha a^2+(p+\frac{1-p}{d})^\alpha b^2]^{\frac{1}{\alpha}-1}$. So $\frac{\partial r}{\partial a}\geq 0$, and $\frac{\partial C_{\alpha}}{\partial a}\geq 0$.
When $\alpha\in (1,2]$,  we can show $\frac{\partial r}{\partial a}\leq 0$, and $\frac{\partial C_{\alpha}}{\partial a}\geq 0$ by a similar way. Therefore $C_{\alpha}$ is an increasing function with respect to $a$ for any $\alpha\in(0,1)\cup (1,2]$.

\textbf{Theorem 4}:  For all n-qubit $X$ states with a  fixed mixedness $M=p$, coherence measures $C_{l_{1}}$,$C_{r}$ and $C_{\alpha}$ will take the same ordering, where  $\alpha\in(0,1)\cup (1,2]$.

According to  Proposition 4, the proof is clear. Theorem 4 gives another sufficient condition that these coherence measures  generate the same ordering for some sets of d-dimensional states.

\section{Dynamics of coherence ordering under Markovian channels}\label{sec:dy}

In this section, we will discuss dynamics of coherence-induced ordering under
Markovian one-qubit channels for single-qubit states with a fixted mixtedness. Here, we only consider Amplitude damping channel and Phase damping channel. We can consider other  Markovian channels by a similar method.

\subsection{ Amplitude damping channel}

Now, we study the dynamics of coherence-induced ordering under the  amplitude damping channel
(ADC), which can be characterized by the Kraus' operators
$K^{AD}_{0}=|0\rangle\langle0|+\sqrt{p}|1\rangle\langle1|I$, $K^{AD}_{1}=\sqrt{p}|0\rangle\langle1|$,
where parameters $p,q \in[0,1]$ and $p+q=1$. Using the amplitude damping channel into the state with the form Eq. (\ref{equa1}), we get

\begin{equation}
\varepsilon(\rho)=\left[ \begin {array}{cc} \frac{1+tn_{z}}{2}+p\frac{1-tn_{z}}{2}&\sqrt{q}t\frac{n_{x}-in_{y}}{2}\\ \noalign{\medskip}\sqrt{q}t\frac{n_{x}+in_{y}}{2}&q\frac{1-tn_{z}}{2}\end {array} \right].\nonumber
\end{equation}
The state $\varepsilon(\rho)$ can be represented by the form Eq. (\ref{equa1}).  The parameters are
$t'=\sqrt{qt^2 (1-n^2_{z})+(p+qn_{z}t)^2}$,
$n'_{x}=\frac{\sqrt{q}n_{x}t}{t'}$,
$n'_{y}=\frac{\sqrt{q}n_{y}t}{t'}$,
$n'_{z}=\frac{p+qn_{z}t}{t'}$.
Substituting these parameters  into  Eq. (\ref{l1norm1}), (\ref{rel1}), (\ref{alpha1}), we obtain

\begin{equation}\label{adclinorm}
C_{l_{1}}(\varepsilon(\rho))=qt\sqrt{1-n_{z}^2}=qC_{l_{1}}(\rho),
\end{equation}

\begin{equation}\label{adcrel}
C_{r}(\varepsilon(\rho))=h(\frac{1+n'_{z}}{2})-h(\frac{1+t'}{2}),
\end{equation}

\begin{equation}\label{adcalpha}
C_{\alpha}(\varepsilon(\rho))=\frac{r^{\alpha}-1}{\alpha-1},
\end{equation}
where
\begin{equation}
\begin{aligned}
r=&[(\frac{1+t'}{2})^\alpha\frac{1+n'_{z}}{2}+(\frac{1-t'}{2})^\alpha\frac{1-n'_{z}}{2}]^\frac{1}{\alpha}
  +[(\frac{1+t'}{2})^\alpha\frac{1-n'_{z}}{2}+(\frac{1-t'}{2})^\alpha\frac{1+n'_{z}}{2}]^\frac{1}{\alpha}.
 \end{aligned}
\end{equation}

In accordance with the Eq. (\ref{adclinorm}), the amplitude
damping channel don't change the coherence ordering induced by the $l_{1}-$ norm of coherence for the single-qubit states.


In the following, we will use the numerical method to discuss dynamics of coherence ordering
 with $C_{r}$ and $C_{\alpha}$ under Markovian channels for single-qubit states.

 We fix the value $p=0.5$, as presented in Fig. 2 and Fig. 3, $C_{r}(\varepsilon(\rho))$ and  $C_{2}(\varepsilon(\rho))$ are increasing functions with respect to $t$ for every fixed  $n_{z}$. By Theorem 2, we  know that amplitude damping channel don't change the coherence-induced ordering by $C_{r}$ or $C_{2}$ with fixed valued $n_{z}$. As presented in Fig. 4, we know $C_{r}(\varepsilon(\rho))$ and  $C_{2}(\varepsilon(\rho))$ aren't monotonic functions with respect to $n_{z}$ for some fixed $t$.  By Theorem 1,
 amplitude damping channel will change the coherence-induced ordering by $C_{r}$ or $C_{2}$ with fixed mixedness.

\begin{center}
\vspace{-0.2cm}
\includegraphics [scale=0.25]{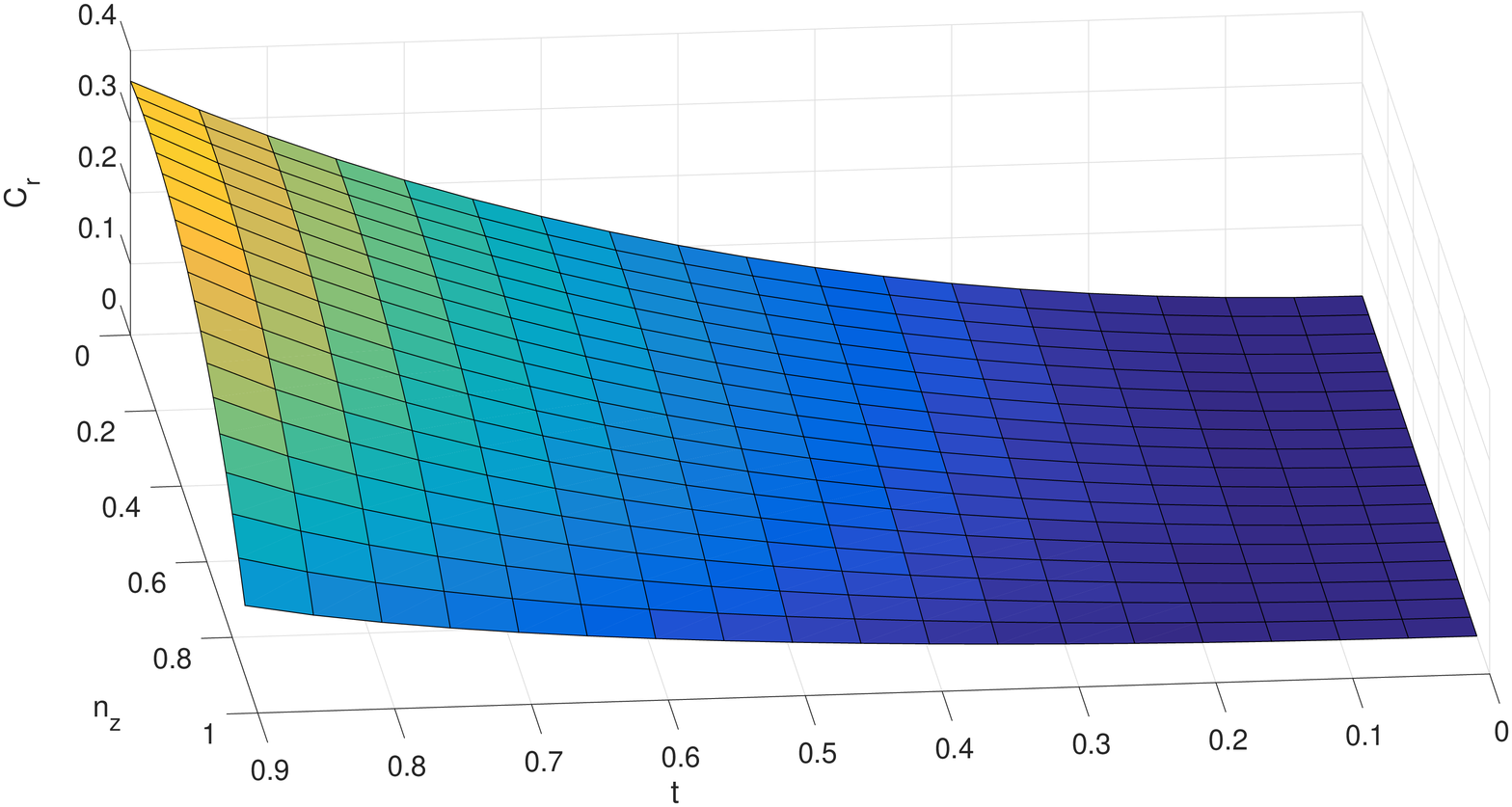}
\vspace{-0.2cm}
\center{Fig. 2.
 For fixed $p=0.5$, the variation of $C_{r}(\varepsilon(\rho))$  with $t$
and $n_{z}$ under phase damping channel.}
\end{center}

\begin{center}
\vspace{-0.2cm}
\includegraphics [scale=0.25]{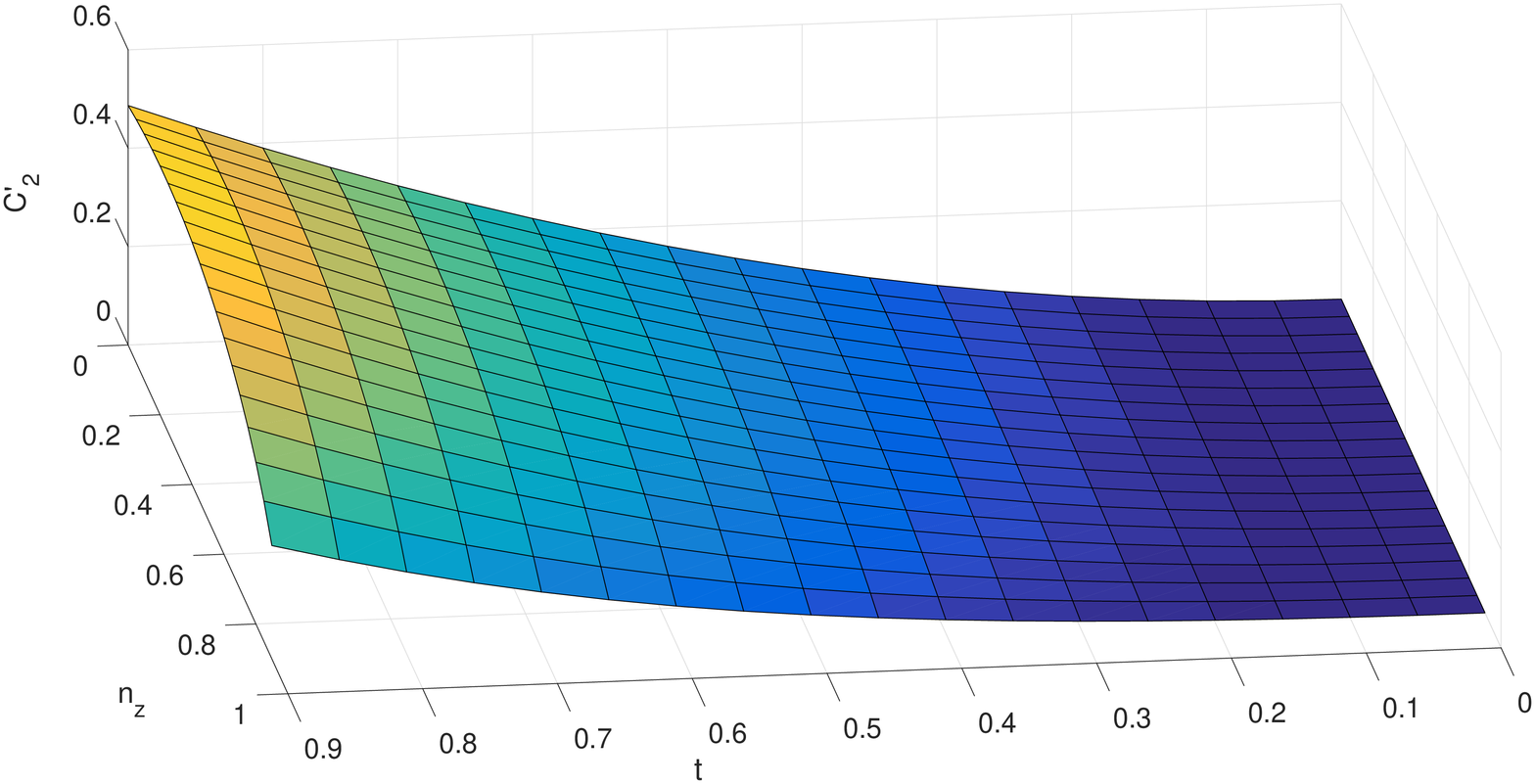}
\vspace{-0.2cm}
\center{Fig. 3.
 For fixed $p=0.5$, the variation of $C_{2}(\varepsilon(\rho))$  with $t$
and $n_{z}$ under phase damping channel.}
\end{center}

\begin{center}
\vspace{-0.2cm}
\includegraphics [scale=0.25]{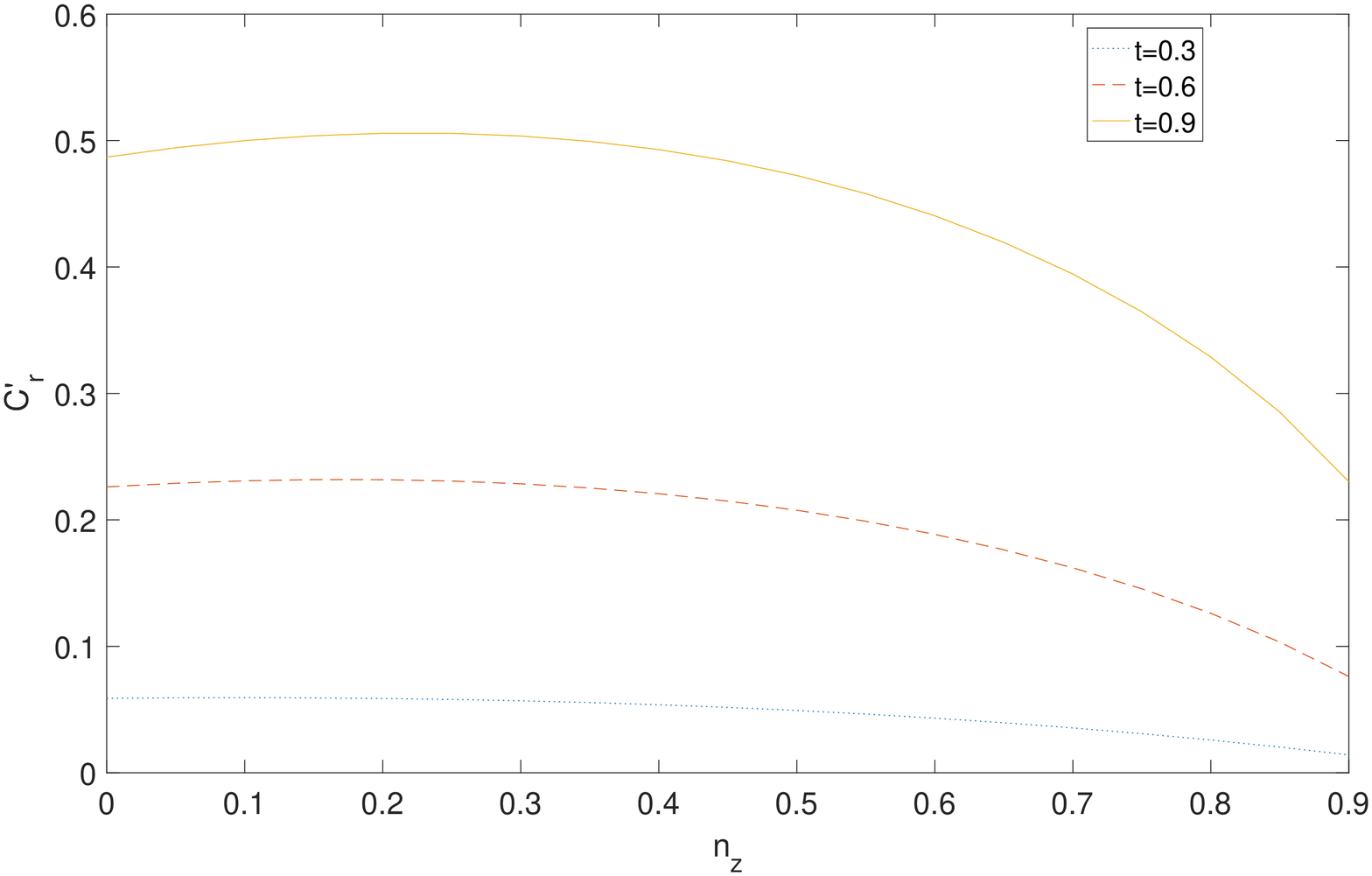}
\vspace{-0.2cm}
\center{Fig. 4.
 For fixed $p=0.5$, the variation of $C_{r}(\varepsilon(\rho))$  with
 $n_{z}$ for fixed $t$ under phase damping channel.}
\end{center}

\begin{center}
\vspace{-0.2cm}
\includegraphics [scale=0.25]{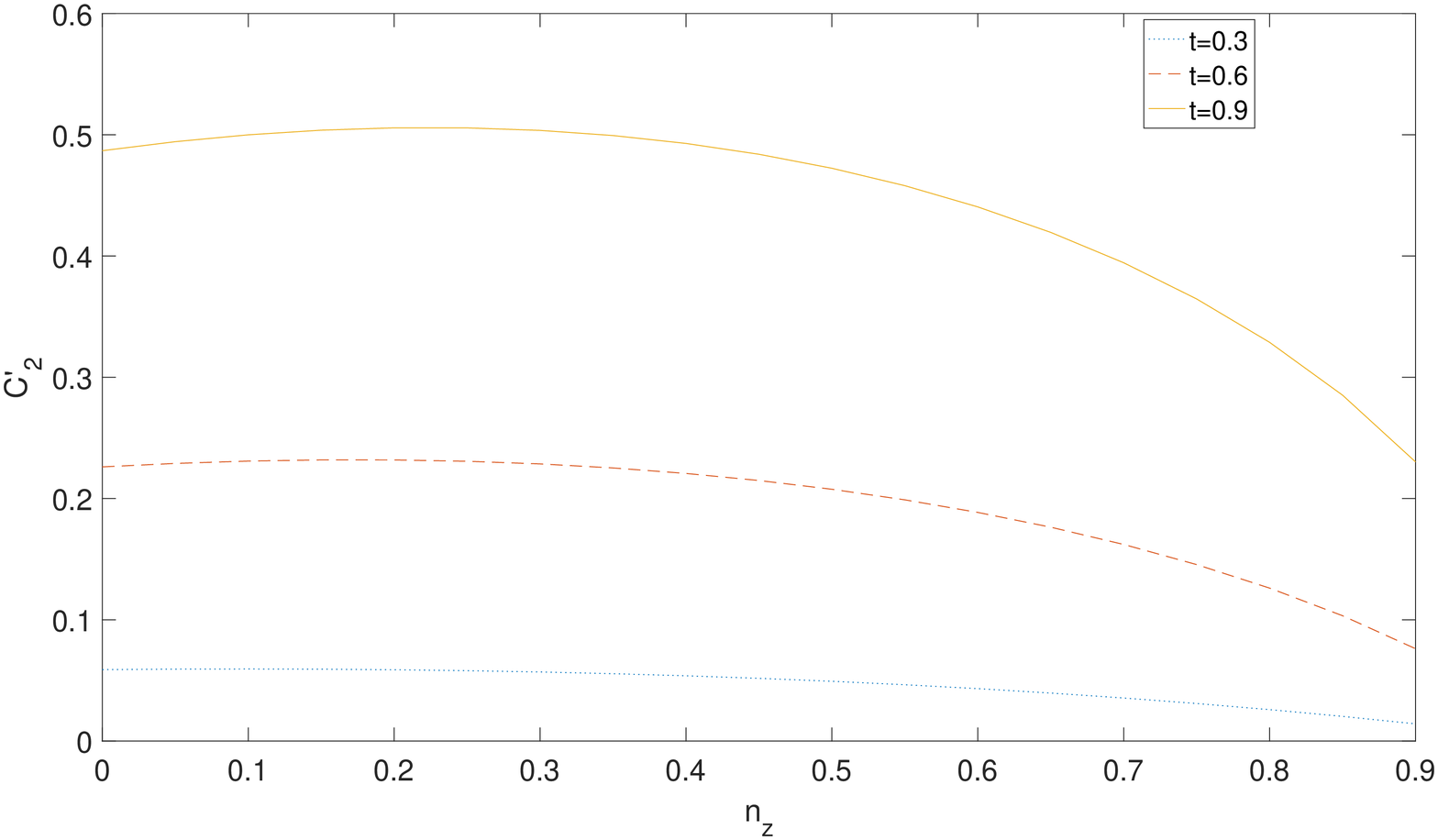}
\vspace{-0.2cm}
\center{Fig. 5.
 For fixed $p=0.5$, the variation of $C_{2}(\varepsilon(\rho))$  with
$n_{z}$ for fixed $t$ under phase damping channel.}
\end{center}

\subsection{ Phase damping channel}

Now, we study the dynamics of of coherence-induced ordering under the phase damping channel
(PDC), which can be characterized by the Kraus' operators
$K^{PD}_{0}=\sqrt{q}I$,$K^{PD}_{1}=\sqrt{p}|0\rangle\langle0|$, $K^{PD}_{2}=\sqrt{p}|1\rangle\langle1|$,
where parameters $p,q \in[0,1]$ and $p+q=1$. Using the phase damping channel into the state with the form Eq. (\ref{equa1}), we get

\begin{equation}
\varepsilon(\rho)=\left[ \begin {array}{cc} \frac{1+tn_{z}}{2}&qt\frac{n_{x}-in_{y}}{2}\\ \noalign{\medskip}qt\frac{n_{x}+in_{y}}{2}&\frac{1-tn_{z}}{2}\end {array} \right].\nonumber
\end{equation}

The state $\varepsilon(\rho)$ can be represented by the form Eq. (\ref{equa1}).  The parameters are
$t'=\sqrt{q^2t^2 +(1-q^2)n_{z}^2t^2}$,
$n'_{x}=\frac{qn_{x}t}{t'}$,
$n'_{y}=\frac{qn_{y}t}{t'}$,
$n'_{z}=\frac{n_{z}t}{t'}$.
Substituting these parameters  into  Eq. (\ref{l1norm1}), (\ref{rel1}), (\ref{alpha1}), we obtain


\begin{equation}
C_{l_{1}}(\varepsilon(\rho))=qt\sqrt{1-n_{z}^2}=qC_{l_{1}}(\rho),
\end{equation}

\begin{equation}
C_{r}(\varepsilon(\rho))=h(\frac{1+n'_{z}}{2})-h(\frac{1+t'}{2}),
\end{equation}

\begin{equation}
C_{\alpha}(\varepsilon(\rho))=\frac{r^{\alpha}-1}{\alpha-1},
\end{equation}
where
\begin{equation}
\begin{aligned}
r=&[(\frac{1+t'}{2})^\alpha\frac{1+n'_{z}}{2}+(\frac{1-t'}{2})^\alpha\frac{1-n'_{z}}{2}]^\frac{1}{\alpha}\\
  +&[(\frac{1+t'}{2})^\alpha\frac{1-n'_{z}}{2}+(\frac{1-t'}{2})^\alpha\frac{1+n'_{z}}{2}]^\frac{1}{\alpha}.
 \end{aligned}
\end{equation}

 Let $p=0.5$, as presented in Fig. 6, and Fig. 7, $C_{r}(\varepsilon(\rho))$ and  $C_{2}(\varepsilon(\rho))$ are increasing functions with respect to $n_{z}$ for every fixed $t$. By Theorem 1, we know that phase damping channel don't change coherence-induced ordering by $C_{r}$ or $C_{2}$ with fixed $t$. Similarly, we know $C_{r}(\varepsilon(\rho))$ and  $C_{2}(\varepsilon(\rho))$ are increasing functions with respect to $t$ for every fixed valued $n_{z}$.
 By Theorem 2, phase damping channel won't change coherence-induced ordering by $C_{r}$ or $C_{2}$ with fixed $n_{z}$.

\begin{center}
\vspace{-0.2cm}
\includegraphics [scale=0.25]{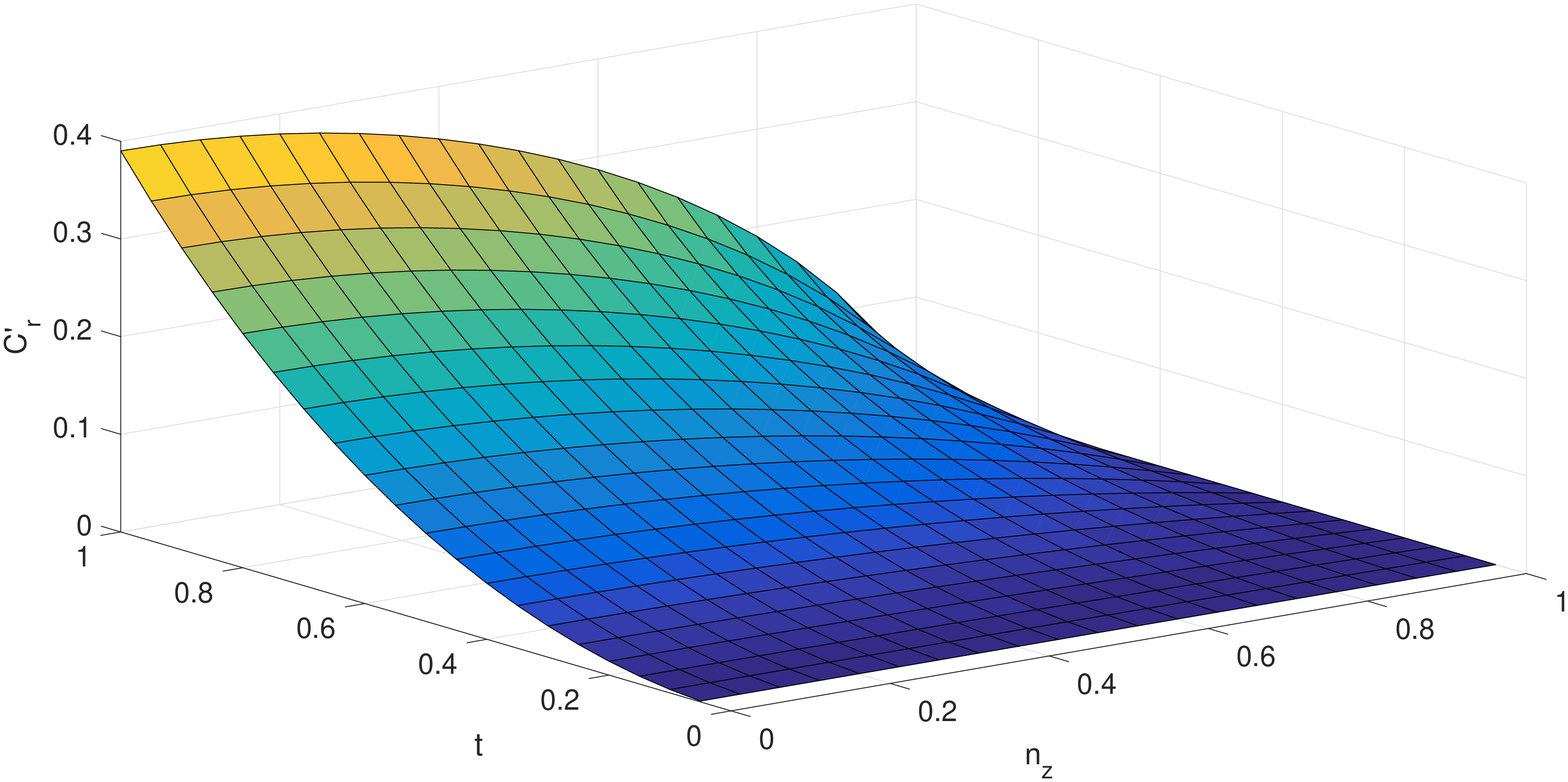}
\vspace{-0.2cm}
\center{Fig. 6.
 For fixed $p=0.5$, the variation of $C_{r}(\varepsilon(\rho))$  with $t$
and $n_{z}$ under phase damping channel.}
\end{center}

\begin{center}
\vspace{-0.2cm}
\includegraphics [scale=0.25]{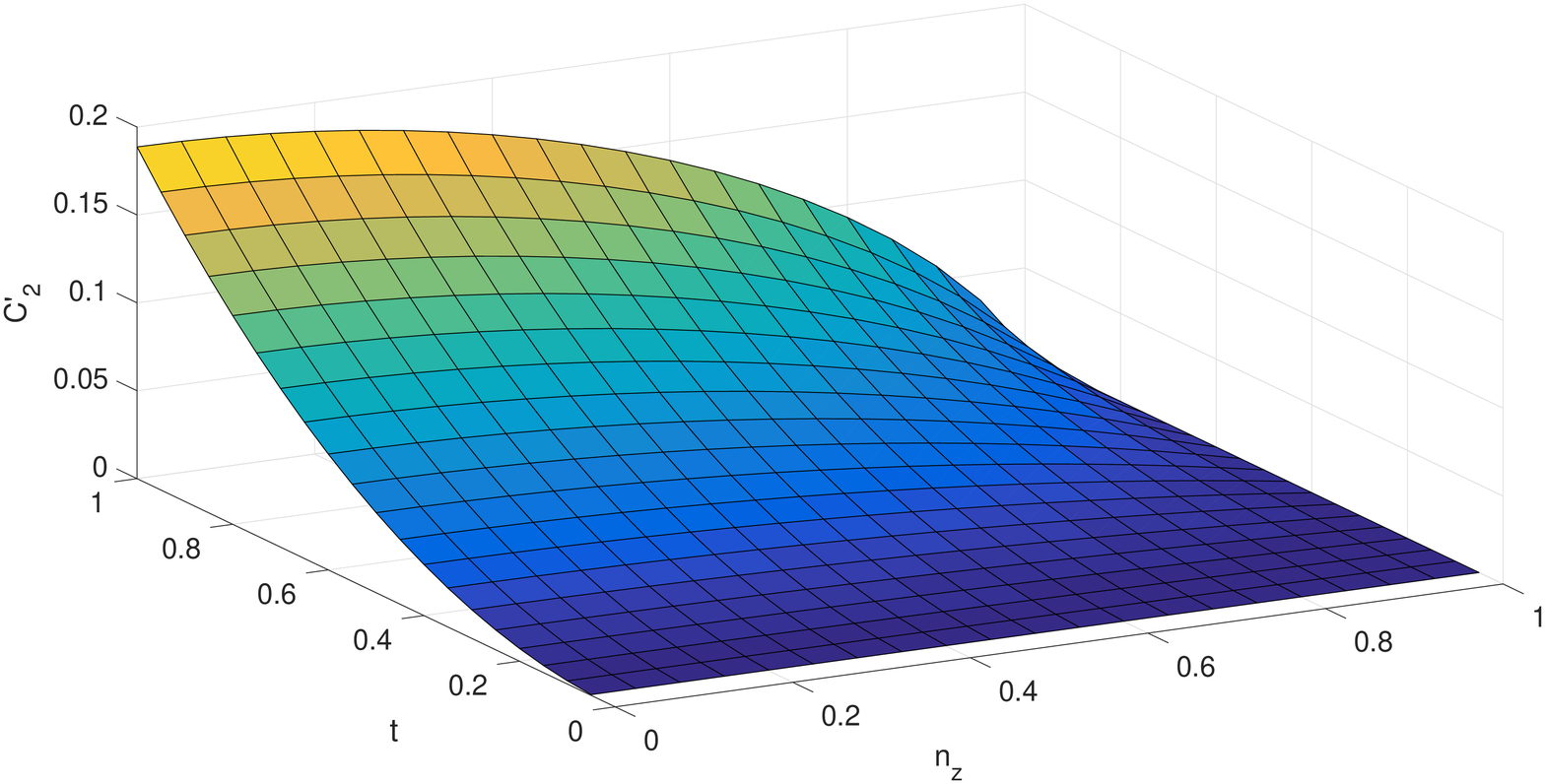}
\vspace{-0.2cm}
\center{Fig. 7.
 For fixed $p=0.5$, the variation of $C_{2}(\varepsilon(\rho))$  with $t$
and $n_{z}$ under phase damping channel.}
\end{center}

\section{Conclusion}\label{sec:conclusion}
In this paper, we studied coherence-induced state ordering with Tsallis relative entropy of coherence, relative entropy of coherence and $l_{1}$ norm of coherence. First, we showed that these three measures give the same ordering for single-qubit states with a fixed mixedness or a fixed length along the direction $\sigma_{z}$.
Second, we considered some special cases of high dimensional states, we showed that these three measures generate the same ordering for the set of high dimensional pure states if any  two states  of the set satisfy majorization relation. Moreover, these three measures generate the same ordering for all $X$ states with a fixed mixedness.  Finally, we discussed dynamics of coherence-induced ordering under Markovian channels. We found phase damping channel don't change the coherence-induced ordering for some single-qubit states with fixed mixedness, but amplitude damping channel change the coherence-induced ordering even though for single-qubit states with fixed mixedness.  We can consider other  Markovian channels by a similar method.

\section{Acknowledgments}
This paper is supported by National Natural Science Foundation of China(Grants No. 11671244, No.11271237),
The Higher School Doctoral Subject Foundation of Ministry of Education of China(Grant No.20130202110001),
and Fundamental Research Funds for the Central Universities( No.2016CBY003).


\begin{thebibliography}{}

\bibitem{Scully97} M. O. Scully and M. S. Zubairy, Quantum Optics (Can-
brudge University Press, Cambridge, 1997).

\bibitem{Nielsen} M. A. Nielsen and I. L. Chuang  {\it Quantum Computation and Quantum Information} (Cambridge: Cambridge Univ. Press) (2000).

\bibitem{Rodr13} C. A. Rodr\'{\i}guez-Rosario, T. Frauenheim, and A. Aspuru-Guzik.: Thermodynamics of quantum coherence. arXiv:1308.1245.


\bibitem{berg14} J ${\AA}$ berg.: Catalytic coherence
 Phys. Rev. Lett. \textbf{113}, 150402 (2014).

\bibitem{Horodecki13} M. Horodecki and J. Oppenheim.: Resource theory of quantum states out of thermal equilibrium.  Nat. Commun. \textbf{4}, 2059 (2013).

\bibitem{Lostaglio15} M. Lostaglio, K. Korzekwa, D. Jennings, and T. Rudolph.: Quantum coherence, time-translation symmetry, and thermodynamics.
Phys. Rev. X \textbf{5}, 021001 (2015).

\bibitem{Naras15} V. Narasimhachar and G. Gour.: Low-temperature thermodynamics with quantum coherence. Nat. Commun. \textbf{6}, 7689 (2015).

\bibitem{berg06}  J. ${\AA}$ berg.: Quantifying Superposition. arXiv:0612146.

\bibitem{Baum14} T. Baumgratz, M. Cramer, and M. B. Plenio.: Quantifying coherence.
 Phys. Rev. Lett.
\textbf{113}, 140401 (2014).

\bibitem{Rastegin16} A .E. Rastegin.: Quantum-coherence quantifiers based on the Tsallis relative ¦Á entropies. Phys. Rev. A \textbf{93}, 032136 (2016).


\bibitem{Swapan16} S. Rana, P. Parashar, and M. Lewenstein.: Trace-distance measure of coherence. Phys. Rev. A \textbf{93}, 012110 (2016).

\bibitem{Yuan15} X. Yuan, H. Zhou, Z. Cao, and X. Ma.: Intrinsic randomness as a measure of quantum coherence. Phys. Rev. A \textbf{92},
022124 (2015).

\bibitem{Shao15} L.-H. Shao, Z. Xi, H. Fan, and Y. Li.: Fidelity and trace-norm distances for quantifying coherence. Phys. Rev. A \textbf{91},
042120 (2015).

\bibitem{Strel15}  A. Streltsov, U. Singh, H. S. Dhar, M. N. Bera, and G.
Adesso.: Measuring quantum coherence with entanglement.
 Phys. Rev. Lett. \textbf{115}, 020403 (2015).

\bibitem{Napoli16}  C. Napoli, T. R. Bromley,  M. Cianciaruso,  M. Piani, N. Johnston, and G. Adesso.: Robustness of coherence: An operational and observable measure of quantum coherence.
  Phys. Rev. Lett. \textbf{116},150502 2016.

\bibitem{Zhang16} Y.-R. Zhang, L.-H. Shao, Y. Li, and H. Fan.: Quantifying coherence in infinite-dimensional systems.
  Phys. Rev. A \textbf{93}, 012334 (2016).

\bibitem{Yu16} X.-D. Yu, D.-J. Zhang, G. F. Xu, D. M. Tong.: An alternative framework for quantifying coherence. arXiv:1606.03181.

\bibitem{Chin17} S. Chin. Generalized Coherence Concurrence and Coherence Number. arXiv:1702.06061.

Rana S, Parashar P, Lewenstein M. Trace-distance measure of coherence[J]. Physical Review A, 2016, 93(1): 012110.


\bibitem{Liu16}  C. L. Liu , X. D. Yu , G. F. Xu , D. M. Tong.: Ordering states with coherence measures. Quantum Information Processing. DOI: 10.1007/s11128-016-1398-5(2016).

\bibitem{Zhangfu16} F. G. Zhang, L. H. Shao, Y. Luo, Y. M. Li. Ordering states with Tsallis relative -entropies of coherence, Quantumn Information Processing. DOI:10.1007/s11128-016-1488-4 (2016)

\bibitem{Yao15} Y. Yao, X. Xiao, L. Ge, and C. P. Sun.: Quantum coherence in multipartite systems. Phys. Rev. A \textbf{92},
022112 (2015).

\bibitem{Du15} S. Du, Z. Bai and Y. Guo.: Conditions for coherence transformations under incoherent operations. Phys. Rev. A \textbf{91}, 052120
(2015).

\bibitem{Cheng15} S. Cheng and M. J. W. Hall.: Complementarity relations for quantum coherence. Phys. Rev. A \textbf{92}, 042101(2015).

\bibitem{Bera15} M. N. Bera, T. Qureshi, M. A. Siddiqui, and A. K. Pati.: Duality of quantum coherence and path distinguishability. Phys. Rev. A \textbf{92}, 012118 (2015).


\bibitem{Xi15} Z. Xi, Y. Li, and H. Fan.: Quantum coherence and correlations in quantum system. Sci. Rep. \textbf{5}, 10922 (2015).

\bibitem{Winter16}  A. Winter and D. Yang.: Operational resource theory of coherence. Phys. Rev. Lett. \textbf{116}, 120404 (2016).

\bibitem{Bromley15} T. R. Bromley, M. Cianciaruso, and G. Adesso.: Frozen quantum coherence. Phys. Rev. Lett. \textbf{114}, 210401 (2015).

 \bibitem{Xu16} J. Xu.:  Quantifying coherence of Gaussian states. Phys. Rev. A \textbf{93}, 032111 (2016).

\bibitem{Yadin15} B. Yadin, J. Ma, D. Girolami, M. Gu, and V. Vedral.: Quantum processes which do not use coherence. arXiv:1512.02085.

\bibitem{Bagan15} E. Bagan, J. A. Bergou, S. S. Cottrell, and M. Hillery.: Relations between coherence and path information.
Phys. Rev. Lett. \textbf{116}, 160406 (2016).

\bibitem{Chitambar16} E. Chitambar, G. Gour.: Are Incoherent Operations Physically Consistent?--A Critical Examination of Incoherent Operations. arXiv:1602.06969.

\bibitem{Peng16}  Y. Peng ,  Y. Jiang, H. Fan,: Maximally coherent states and coherence-preserving operations. Phys. Rev. A \textbf{93} 032326(2016).


\bibitem{Bran15} G. S. L. F. Brand$\tilde{a}$o, G. Gour.: Reversible framework for quantum resource theories. Phys. Rev. Lett. \textbf{115}, 070503 (2015).

 \bibitem{Liu17}  Z. W. Liu , X. Hu, S. Lloyd. Resource Destroying Maps[J]. Phys. Rev. Lett.\textbf{118}, 060502 (2017).
 \bibitem{Tan17} K. C. Tan, T. Volkoff, H. Kwon, H. Jeong. Quantifying the Coherence Between Coherent States. arXiv:1703.01067.
\bibitem{Hu17} M. L. Hu, X. Hu, Y. Peng, Y. R. Zhang, H. Fan. Quantum coherence and quantum correlations. arXiv:1703.01852.






















\bibitem{Furuichi04} S. Furuichi , K. Yanagi, K. Kuriyama.: Fundamental properties of Tsallis relative entropy. J. Math. Phys. \textbf{45}, 4868 (2004).

\bibitem{Hiai11} F. Hiai,  M. Mosonyi, D. Petz.: Reversibility conditions for quantum operations
. C.B\'{e}ny. Rev. Math. Phy. \textbf{23}, 691 (2011).


\bibitem{Luo16} Y. Luo , T. Tian, L.-H. Shao, Y. Li.:General Monogamy of Tsallis-q Entropy Entanglement in Multiqubit Systems. Phys. Rev. A 93, 062340 (2016).

\bibitem{Peters04} N.A. Peters, T.-C. Wei, P.G. Kwiat, Phys. Rev. A 70 (2004) 052309.

\bibitem{Mar79} A. W. Marshall, I. Olkin, B. C. Arnold.  Inequalities: theory of majorization and its applications[M]. New York: Academic press, 1979.




\end{thebibliography}
\end{document}